\documentclass[twocolumn,english,prl,superscriptaddress,floatfix]{revtex4}
\usepackage[T1]{fontenc}
\usepackage[latin9]{inputenc}
\usepackage{amsbsy}
\usepackage{graphicx}
\usepackage{amssymb}

\makeatletter

\usepackage{amsbsy}

\makeatletter

\usepackage{amsbsy}

\makeatletter

\usepackage{bm}

\usepackage{babel}


\makeatother

\usepackage{babel}

\makeatother

\usepackage{babel}

\makeatother

\usepackage{babel}

\begin{document}

\title{Anisotropic in-plane resistivity in the nematic phase of the iron
pnictides}

\author{Rafael M. Fernandes}

\altaffiliation{Present address: Department of Physics, Columbia University, New York, New York 10027, USA}

\affiliation{Ames Laboratory and Department of Physics and Astronomy, Iowa State
Univ., Ames, IA 50011, USA}

\author{Elihu Abrahams}

\affiliation{Department of Physics and Astronomy, University of California Los
Angeles, Los Angeles, CA 90095, USA}

\author{Jörg Schmalian}

\altaffiliation{Present address: Karlsruhe Institute of Technology, Institute for Theory of Condensed Matter, D-76131 Karlsruhe, Germany}

\affiliation{Ames Laboratory and Department of Physics and Astronomy, Iowa State
Univ., Ames, IA 50011, USA}

\date{\today }

\begin{abstract}
We show that the interference between scattering by impurities and
by critical spin fluctuations gives rise to anisotropic transport
in the Ising-nematic state of the iron pnictides. The effect is closely
related to the non-Fermi liquid behavior of the resistivity near an
antiferromagnetic quantum critical point. Our theory not only explains
the observed sign of the resistivity anisotropy $\Delta\rho$ in electron
doped systems, but also predicts a sign change of $\Delta\rho$ upon
sufficient hole doping. Furthermore, our model naturally addresses
the changes in $\Delta\rho$ upon sample annealing and alkaline-earth
substitution. 
\end{abstract}
\maketitle
In many materials, anisotropic properties are related to the underlying
crystalline structure. However, when correlations are present, the
electronic states can themselves become anisotropic \cite{Kivelson98}.
Recent transport measurements in detwinned crystals of the iron pnictide
compounds $\mathrm{\mathit{AE}Fe_{2}As_{2}}$ ($\mathit{AE}=\mathrm{Ba}$,
$\mathrm{Ca}$, $\mathrm{Sr}$) found an in-plane anisotropy that
cannot be attributed only to lattice distortions, unveiling an anisotropic
electronic state \cite{Chu10,Tanatar10}. Its existence is also supported
by the observations of an orbital polarization $\mathcal{O}=n_{xz}-n_{yz}$
of $d_{xz}$ and $d_{yz}$ Fe states in angle-resolved photoemission
spectroscopy (ARPES) \cite{Shen11}, local anisotropies in scanning
tunneling microscopy \cite{Davis10}, and anisotropies in the optical
spectrum \cite{Duzsa11,Lucarelli11,Uchida11}.

One candidate for such unconventional electronic state is the Ising-nematic
order, which emerges from the combination of magnetic fluctuations
and frustration \cite{Si08,Yildirim08,Fang08,Xu08,FernandesPRL10}.
In a strong-coupling approach, frustration is promoted by the competing
$J_{1}-J_{2}$ exchange interactions, while in an weak-coupling approach,
it follows from the degeneracy of the magnetic ground state due to
the nesting properties of the Fermi surface (FS). Indeed, the iron
pnictides support two magnetic instabilities with order parameters
$\boldsymbol{\Delta}_{1}$, $\boldsymbol{\Delta}_{2}$ corresponding
to the in-plane ordering vectors $\mathbf{Q}_{1}=\left(\pi,0\right)$,
$\mathbf{Q}_{2}=\left(0,\pi\right)$, respectively. The electronic
structure \cite{Eremin10,Antropov11}, as well as the coupling to
the lattice \cite{FernandesPRL10}, give rise to the coupling $g\Delta_{1}^{2}\Delta_{2}^{2}$
in the free energy, with $g>0$. As a result, a discrete Ising-nematic
degree of freedom $\varphi\propto\left\langle \Delta_{1}^{2}\right\rangle -\left\langle \Delta_{2}^{2}\right\rangle $
emerges, which labels the two degenerate states $\Delta_{1}\neq0$,
$\Delta_{2}=0$ and $\Delta_{1}=0$, $\Delta_{2}\neq0$. Most importantly,
$\varphi$ is able to order at a temperature $T_{s}$ above the onset
of antiferromagnetism (AFM) at $T_{N}$ \cite{chandra}, breaking
the tetragonal symmetry already in the paramagnetic (PM) phase - hence
the term nematic.

By symmetry, Ising-nematic order induces the orbital polarization
$\mathcal{O}$ and a shear distortion $\varepsilon_{s}$, driving
a structural transition from the tetragonal (Tet) to the orthorhombic
(Ort) phase at $T_{s}$. It explains why $T_{s}$ and $T_{N}$ track
each other closely across every pnictide phase diagram, even inside
the superconducting (SC) dome \cite{Nandi09,FernandesPRB10}. Recently,
it has been shown that nematic fluctuations can\textbf{ }explain the
dramatic softening of the lattice observed experimentally in the Tet
phase \cite{FernandesPRL10}. It remains an important issue whether
this scenario is also able to address the observed in-plane resistivity
anisotropy $\Delta\rho\equiv\rho_{b}-\rho_{a}$, where $b$ ($a$)
refers to the direction with shorter (longer) lattice constant of
the Ort state. The induced orbital polarization $\mathcal{O}$ has
been alternatively proposed to explain such anisotropy, but different
authors disagree on the suitability of this proposal \cite{Chen10,Calderon10}.

\begin{figure}
\begin{centering}
\includegraphics[width=0.9\columnwidth]{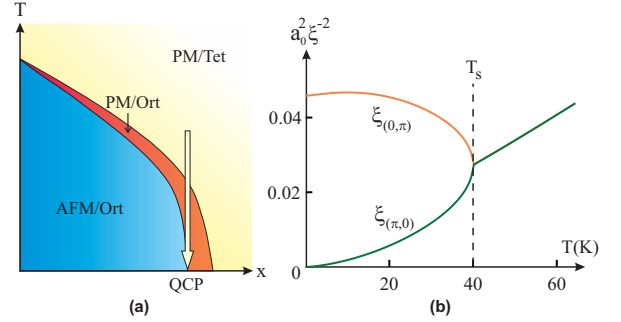} 
\par\end{centering}

\caption{\textbf{(a)} Schematic temperature-doping ($T,x$) phase diagram in
the absence of SC. The yellow region corresponds to enhanced nematic
fluctuations, as seen by elastic modulus measurements \cite{FernandesPRL10}.
\textbf{(b)} Temperature dependence of the correlation lengths $\xi$
associated with the $\left(\pi,0\right)$ and $\left(0,\pi\right)$
magnetic instabilities above the magnetic QCP.}

\end{figure}

In this paper we demonstrate that the observed anisotropic transport
\emph{above} $T_{N}$ can be understood within the Ising-nematic picture.
Our model explains in a unified way several experimental observations,
such as the sign of the anisotropy \cite{Chu10,Tanatar10} and the
changes in $\Delta\rho$ upon sample annealing \cite{Uchida11} or
alkaline-earth substitution \cite{Blomberg11}. We also predict a
sign change of $\Delta\rho$ for hole-doped systems, thus addressing
the peculiar observations in $K$-doped compounds \cite{Ying10,Tanatar_private}.
Our results support the view that nematic order is the driving force
for the structural transition and the orbital polarization in the
pnictides. Note that our theory does not address the anisotropy observed
\emph{below} $T_{N}$, in the AFM state. However, in all observations,
the sign of $\Delta\rho$ is already set above $T_{N}$, where the
slope $d\left(\Delta\rho\right)/dT$ is the largest \cite{Chu10,Tanatar10}.
Concentrating on the regime above $T_{N}$ allows for a more transparent
and less model-dependent understanding of the resistivity anisotropy.

Below $T_{s}$, the onset of a finite Ising-nematic order parameter
$\varphi\neq0$ gives rise to two sources of anisotropy: in the electronic
structure itself and in the spectrum of magnetic fluctuations. The
former is rather weak, due to the smallness of the shear distortion
$\varepsilon_{s}=\left(a-b\right)/\left(a+b\right)$ - for instance,
in optimally doped $\mathrm{Ba\left(Fe_{1-x}Co_{x}\right)_{2}As_{2}}$,
$\varepsilon_{s}\approx5\times10^{-4}$ \cite{Shen11,Nandi09}. On
the other hand, the latter is strong, since only fluctuations associated
with one of the magnetic instabilities diverge. The impact of this
anisotropy on the transport depends on the contribution of the scattering
of electrons by spin fluctuations around the hot spots of the Fermi
surface, defined via $\varepsilon_{\lambda,\mathbf{k}}=\varepsilon_{\lambda',\mathbf{k}+\mathbf{Q}_{i}}$,
with $\varepsilon_{\lambda,\mathbf{k}}$ denoting the dispersion of
band $\lambda$. In ultra-clean systems, this scattering channel is
short-circuited by the contribution from the other regions of the
FS \cite{Hlubina95}. However, in the presence of significant impurity
scattering, the scattering by spin fluctuations becomes important,
leading to non-Fermi liquid behavior \cite{Rosch99} and, in addition,
to anisotropic transport. To see this, we follow Rosch \cite{Rosch99}
and solve the Boltzmann equation (BE) for scattering by impurities
and spin fluctuations, which yields features not captured by the relaxation
time approximation \cite{Kemper11}. In the regime where impurity
scattering dominates, we obtain the resistivity along the $\alpha$
($\alpha=x,y,z$) direction:

\begin{equation}
\rho_{\alpha\alpha}-\rho_{\mathrm{imp}}=\sum_{\mathbf{k}\mathbf{k}^{\prime},\lambda\lambda^{\prime}}\left(\Phi_{\mathbf{k},\lambda}^{\alpha}-\Phi_{\mathbf{k}^{\prime},\lambda^{\prime}}^{\alpha}\right)^{2}\mathcal{D}_{\mathbf{kk}^{\prime}}^{\lambda\lambda^{\prime}}\label{Boltzmann}\end{equation}
 with $\mathcal{D}_{\mathbf{kk}^{\prime}}^{\lambda\lambda^{\prime}}\equiv f_{\mathbf{k}^{\prime},\lambda^{\prime}}^{0}\left(1-f_{\mathbf{k},\lambda}^{0}\right)t_{\mathrm{sf},\mathbf{kk}^{\prime}}^{\lambda\lambda^{\prime}}$.
Here, $\mathbf{k}$ is the electron momentum, $\lambda$ is the band
index, $f_{\mathbf{k},\lambda}^{0}$ is the Fermi-Dirac distribution
function, and $\rho_{\mathrm{imp}}$ is the (isotropic) residual resistivity.
$t_{\mathrm{sf},\mathbf{k,k}^{\prime}}^{\lambda\lambda^{\prime}}=\sum_{i=1}^{2}g_{\mathrm{sf,}i}^{\lambda\lambda^{\prime}}n\left(\omega\right)\mathrm{Im}\chi_{i}\left(\mathbf{q},\omega\right)$
is the spin-fluctuations collision term, where $\omega=\varepsilon_{\lambda,\mathbf{k}}-\varepsilon_{\lambda^{\prime},\mathbf{k}^{\prime}}$
and $\mathbf{q=k-k}^{\prime}$ are, respectively, the transferred
energy and momentum, and $\chi_{i}\left(\mathbf{q},\omega\right)$
is the dynamic susceptibility associated with the ordering vector
$\mathbf{Q}_{i}$. $g_{\mathrm{sf,}i}^{\lambda\lambda^{\prime}}$
is the scattering amplitude and $\Phi_{\mathbf{k},\lambda}^{\alpha}$
is the deviation of the electron distribution function from equilibrium
due to scattering by impurities only, as a consequence of an electric
field applied along the $\alpha$ direction. It only depends on the
impurity scattering potential $V_{\mathrm{imp}}^{\lambda\lambda^{\prime}}$
and on the shape of the FS (see supplementary material for more details).

\begin{figure}
\begin{centering}
\includegraphics[width=0.7\columnwidth]{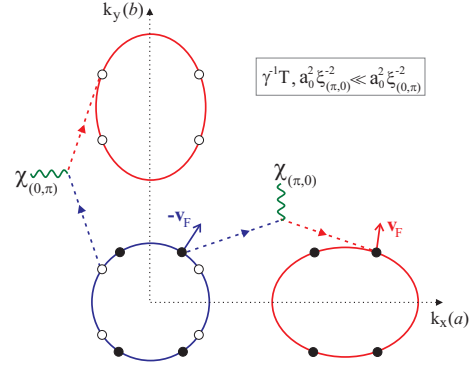} 
\par\end{centering}

\caption{Electrons at the active and passive hot spots (full and empty circles,
respectively) are scattered with different amplitudes by spin fluctuations
associated with the $\left(\pi,0\right)$ and $\left(0,\pi\right)$
ordering vectors. The projection of the hot spots Fermi velocities
(arrows) along the $\left(a,b\right)$ axes determine the sign of
the low-temperature resistivity anisotropy. }

\end{figure}

The resistivity anisotropy data in $\mathrm{Ba\left(Fe_{1-x}Co_{x}\right)_{2}As_{2}}$
show that the temperature-dependent contribution is smaller than the
residual resistivity, particularly close to optimal doping \cite{Chu10}.
Therefore, we can calculate $\rho_{\alpha\alpha}$ using Eq.\ (\ref{Boltzmann}),
whose inputs are the band dispersions $\varepsilon_{\lambda,\mathbf{k}}$,
which determine the impurity-only solution $\Phi_{\mathbf{k},\lambda}^{\alpha}$,
and the dynamic susceptibilities $\chi_{i}\left(\mathbf{q},\omega\right)$,
which determine the spin-fluctuation scattering amplitude $t_{\mathrm{sf},\mathbf{k,k}^{\prime}}^{\lambda\lambda^{\prime}}$.
In the Ising-nematic phase, the latter are:

\begin{equation}
\chi_{i}\left(\mathbf{q},\omega\right)=\frac{C_{0}}{r_{i}\left(\mathbf{q}+\mathbf{Q}_{i}\right)\mp\varphi-i\omega/\gamma},\label{chi}\end{equation}
 where the upper (lower) signs refers to $\mathbf{Q}_{1}$ (\textbf{$\mathbf{Q}_{2}$}),
$\gamma$ is the Landau damping parameter, and $r_{i}\left(\mathbf{q}+\mathbf{Q}_{i}\right)=a_{0}^{2}\xi^{-2}+a_{0}^{2}q_{x}^{2}\left(1\pm\eta\right)+a_{0}^{2}q_{y}^{2}\left(1\mp\eta\right)+\eta_{z}^{2}\cos^{2}\left(\frac{q_{z}c_{0}}{2}\right)$.
This phenomenological form follows naturally from the electronic structure
of the iron pnictides \cite{Zhang10}, and is confirmed by inelastic
neutron scattering experiments \cite{Diallo10}. Here, we introduced
the in-plane (out-of-plane) tetragonal lattice constant $a_{0}$ ($c_{0}$),
the magnetic correlation length $\xi$, the in-plane momentum anisotropy
$\eta$, and the out-of-plane anisotropy $\eta_{z}$. The temperature
dependence of $\xi$ and $\varphi$ are calculated within a self-consistent
mean-field approach \cite{Fang08,Xu08}, yielding the equations:

\begin{eqnarray}
1 & = & \frac{T}{2}\sum_{\mathbf{q},\omega_{n}}\left[\chi_{1}\left(\mathbf{q},\omega_{n}\right)+\chi_{2}\left(\mathbf{q},\omega_{n}\right)\right]\nonumber \\
\frac{\varphi}{g} & = & \frac{T}{2}\sum_{\mathbf{q},\omega_{n}}\left[\chi_{1}\left(\mathbf{q},\omega_{n}\right)-\chi_{2}\left(\mathbf{q},\omega_{n}\right)\right]+h_{\mathrm{strain}},\label{self_consistent}\end{eqnarray}
 where $\omega_{n}$ are bosonic Matsubara frequencies and $h_{\mathrm{strain}}\propto P$,
with $P$ denoting the external strain applied to detwin the samples.
For $h_{\mathrm{strain}}=0$, Eqs. \ref{self_consistent} have a non-trivial
solution with $\varphi\neq0$ below a temperature $T_{s}$, describing
the PM-Ort (Ising-nematic) phase. This model is similar to the one
used in Ref.\ \cite{FernandesPRL10}, which successfully describes
the lattice softening above $T_{s}$. Since $\xi_{i}^{-2}=\xi^{-2}\mp\varphi$,
when $\varphi>0$ (i.e. $a\parallel x$ and $b\parallel y$) the correlation
length $\xi_{\left(\pi,0\right)}$ associated with the $\mathbf{Q}_{1}$
ordering vector lengthens while $\xi_{\left(0,\pi\right)}$, associated
with $\mathbf{Q}_{2}$, shortens (see Fig.\ 1). In the figure, we
present results for an AFM quantum critical point (QCP), $\xi_{\left(\pi,0\right)}^{-1}\left(T=0\right)=0$,
where fluctuations are strongest.

Substituting the susceptibilities given by (\ref{chi}) in the BE
solution (\ref{Boltzmann}), we find that the low-$T$ transport is
dominated by the subset of hot spots connected by the soft ordering
vector ($\mathbf{Q}_{1}$ for $\varphi>0$), $\varepsilon_{\lambda,\mathbf{k}_{hs}}=\varepsilon_{\lambda^{\prime},\mathbf{k}_{hs}+\mathbf{Q}_{1}}$,
while contributions due to $\mathbf{Q}_{2}=\left(0,\pi\right)$ are
negligible. It follows that

\begin{equation}
\rho_{\alpha\alpha}-\rho_{\mathrm{imp}}=\kappa T^{\zeta}\left(\Phi_{\mathbf{k}_{hs},\lambda}^{\alpha}-\Phi_{\mathbf{k}_{hs}+\mathbf{Q}_{1},\lambda^{\prime}}^{\alpha}\right)^{2}\label{anisotropy}\end{equation}
 where $\kappa>0$ is a constant. The exponent $\zeta$ determines
the $T$-dependence of the resistivity and is given by $\zeta=3/2$
($\zeta=1$) for $T\lesssim\Lambda\,\eta_{z}^{2}$ ($T\gtrsim\Lambda\,\eta_{z}^{2}$),
depending on whether the spin fluctuation spectrum is two- or three
dimensional ($\Lambda$ is the energy cutoff). While the experiments
of Ref.\ \cite{Kasahara10} on $\mathrm{BaFe_{2}\left(As_{1-x}P_{x}\right)_{2}}$
materials support $\zeta=1$, Eq.\ (\ref{anisotropy}) is always
dominant when compared to the Fermi-liquid contribution $\rho-\rho_{\mathrm{imp}}\propto T^{2}$.
The distinction into active and passive hot spots is not restricted
to the AFM QCP and persists everywhere along the finite temperature
phase boundary where soft and hard magnetic fluctuations exist. Thus,
the Ising-nematic order causes an anisotropy in the spectrum of magnetic
fluctuations, which induces an anisotropy in the resistivity via scattering
by spin fluctuations (see Fig.\ 2). To determine the amplitude and
sign of $\Delta\rho$, all that is left is to calculate $\Phi_{\mathbf{k},\lambda}^{\alpha}$,
i.e. the solution of the BE with impurities only, which depends solely
on the band dispersions $\varepsilon_{\lambda,\mathbf{k}}$.

To gain analytical insight, we introduce a simplified model with a
circular hole pocket at the center of the Brillouin zone and elliptical
electron pockets displaced from the center by $\mathbf{Q}_{1}$ and
$\mathbf{Q}_{2}$ \cite{FernandesPRB10,Eremin10}, $\varepsilon_{1,\mathbf{k}}=\varepsilon_{0}-k^{2}/2m-\mu$,
$\varepsilon_{2,\mathbf{k}+\mathbf{Q}_{1}}=-\varepsilon_{0}+k_{x}^{2}/2m_{x}+k_{y}^{2}/2m_{y}-\mu$,
and $\varepsilon_{3,\mathbf{k}+\mathbf{Q}_{2}}=-\varepsilon_{0}+k_{x}^{2}/2m_{y}+k_{y}^{2}/2m_{x}-\mu$.
Here, $\varepsilon_{0}$ is the shift in the band energies, $m_{i}$
are the band masses, and $\mu$, the chemical potential. The calculation
of $\Phi_{\mathbf{k},\lambda}^{\alpha}$ for this case is straightforward
and is presented in the supplementary material. In particular, we
find that the sign and amplitude of the anisotropy depends on the
projections along $k_{x}$ (parallel to $a$) and $k_{y}$ (parallel
to $b$) of the Fermi velocities at the hot spots $\varepsilon_{1,\mathbf{k}_{hs}}=\varepsilon_{2,\mathbf{k}_{hs}+\mathbf{Q}_{1}}$
(see Fig.\ 2). When the hot spots are close to a certain axis, there
is stronger scattering by spin fluctuations when the electrons move
parallel to this axis, and the resistivity is larger along this direction.
Interestingly, ARPES measurements in the optimally doped $\mathrm{Ba\left(Fe_{1-x}Co_{x}\right)_{2}As_{2}}$,
which is close to a possible AFM QCP covered by the SC dome \cite{Abrahams09},
reveal the presence of hot spots close to the $k_{y}$ ($b$) axis
\cite{Liu10}. According to our model, this implies $\rho_{b}>\rho_{a}$,
in agreement with transport measurements \cite{Chu10}.

\begin{figure}
\begin{centering}
\includegraphics[width=0.8\columnwidth]{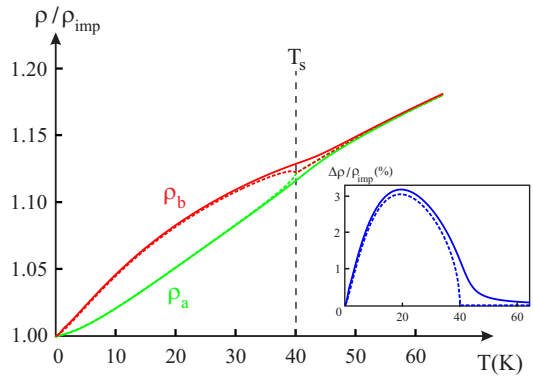} 
\par\end{centering}

\caption{Temperature dependence, above the magnetic QCP, of the resistivities
along $a$ ($\rho_{a}$, green curve) and $b$ ($\rho_{b}$, red curve),
in the absence (dashed lines) and presence (full lines) of external
strain. The inset shows the resistivity anisotropy $\Delta\rho=\rho_{b}-\rho_{a}$.}

\end{figure}

To go beyond the low-$T$ limit in Eq.\ (\ref{anisotropy}), we numerically
calculated $\Delta\rho$ from Eq.\ (\ref{Boltzmann}) for fixed band
structure parameters, $m_{x}=1.25m$, $m_{y}=0.83m$, and $\mu=0.05\varepsilon_{0}$.
For the dynamic susceptibility in Eq.\ (\ref{chi}), based on previous
works \cite{FernandesPRL10} and neutron scattering data \cite{Diallo10},
we used $g/C_{0}\approx2\times10^{-2}$, $\eta_{z}\approx0.1$, $\eta\approx0.5$,
$\gamma\approx350$ meV, and $C_{0}\approx1$ meV, yielding $T_{s}=40$
K and $T_{N}=0$. We also considered $g_{\mathrm{sf,}i}^{\lambda\lambda^{\prime}}\approx2V_{\mathrm{imp}}^{\lambda\lambda^{\prime}}$
and expressed the results in terms of the residual resistivity $\rho_{\mathrm{imp}}$.

The behavior of $\rho_{a}$ and $\rho_{b}$ as function of temperature
is shown in Fig.\ 3, where we considered $\varphi>0$, i.e. $x$
and $y$ parallel to $a$ and $b$, respectively. In accordance to
our discussion of the low-$T$ limit, we find $\rho_{b}>\rho_{a}$
for any temperature below $T_{s}$, a behavior that lingers even when
one moves away from the QCP. As expected, we also find that $\Delta\rho\sim\varphi$
close to $T_{s}$ \cite{Westfahl98}. Note that the sign of $\Delta\rho$
does not depend on the sign of the in-plane momentum anisotropy $\eta$
of $\chi\left(\mathbf{q},\omega\right)$ in Eq.\ (\ref{chi}). As
$T\rightarrow0$, spin fluctuations are suppressed and $\Delta\rho\rightarrow0$.
In the actual data of Ref.\ \cite{Chu10}, these curves are cut off
by the onset of SC, which is not considered in our model.

To make the comparison with experiments more realistic, we also included
the effects of a small strain $P=0.1$ MPa, which is applied to detwin
the sample. This strain breaks the tetragonal symmetry at any temperature,
and there is not a well-defined structural transition, since $\varphi\sim h_{\mathrm{strain}}^{1/\delta}$
is finite at $T_{s}^{\mathrm{twin}}$, where $\delta$ is the corresponding
Ising critical exponent. Our calculations show that the AFM transition
is practically unchanged for small strain, in agreement with the observations
in Ref.\ \cite{Chu10}. If the structural transition in the twin
sample is strongly first order such that $h_{\mathrm{strain}}^{1/\delta}$
is small compared to the jump of $\varphi$ at $T_{s}^{\mathrm{twin}}$,
the tail in $\Delta\rho$ above $T_{s}^{\mathrm{twin}}$ becomes negligible.
This is the case in the parent compounds with $\mathrm{Ba}$ replaced
by $\mathrm{Ca}$ or $\mathrm{Sr}$, as observed by Ref.\ \cite{Blomberg11}.

Our model also explains recent experiments that found a suppression
of $\Delta\rho$ after annealing the sample \cite{Uchida11}. As we
mentioned earlier, in the ultra-clean limit the contribution of the
hot spots to the resistivity is short-circuited by the other regions
of the FS \cite{Hlubina95}. Since the anisotropy is governed by scattering
processes near the hot spots, $\Delta\rho$ becomes smaller for cleaner
samples. Our results can also be rationalized in terms of a $T$-dependent
anisotropic dressing of the impurity scattering by spin fluctuations.
Via this many-body mechanism, an $s$-wave scattering center acquires
effectively an anisotropic cross-section, breaking the $C_{4}$ symmetry.

\begin{figure}
\begin{centering}
\includegraphics[width=0.75\columnwidth]{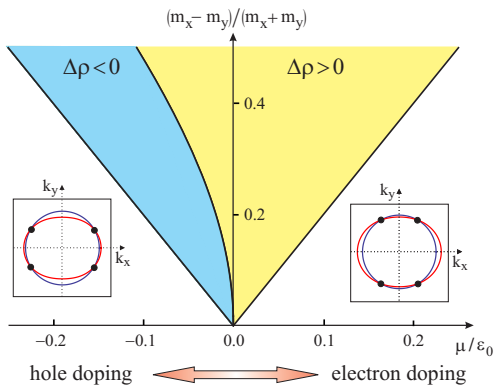} 
\par\end{centering}

\caption{Sign of the low-temperature resistivity anisotropy $\Delta\rho=\rho_{b}-\rho_{a}$
as function of the ellipticity $\frac{m_{x}-m_{y}}{m_{x}+m_{y}}$
of the electron pockets and the chemical potential $\mu$. Note the
asymmetry between electron and hole doping.}

\end{figure}

Finally, we discuss a general prediction of our model. When doping
is introduced and the chemical potential shifts, the positions of
the hot spots change accordingly. Without relying on details of $\varepsilon_{\lambda,\mathbf{k}}$,
one expects that by hole-doping the samples (i.e. decreasing $\mu$
towards negative values), the area of the hole pocket increases, whereas
the area of the electron pocket decreases. Consequently, active hot
spots, which initially are close to the $k_{y}$ ($b$) axis, will
move towards the $k_{x}$ ($a$) axis, and $\Delta\rho$ will eventually
change its sign. This is illustrated in Fig.\ 4, where we present
a phase diagram for the sign of $\Delta\rho$ for different values
of the chemical potential $\mu$ and the ellipticity of the electron
pockets $\propto m_{x}-m_{y}$, keeping the other parameters unchanged.
It is interesting to note that measurements of $\Delta\rho$ in hole-doped
samples $\mathrm{\left(Ba_{1-x}K_{x}\right)Fe_{2}As_{2}}$ found a
vanishingly small resistivity anisotropy \cite{Ying10}, as expected
near the region of the phase diagram of Fig.\ 4 where $\Delta\rho$
changes sign. Most interestingly, very recent data on samples with
higher doping revealed a negative $\Delta\rho$, in accordance to
our prediction \cite{Tanatar_private}.

In our approach, the resistivity anisotropy above $T_{N}$ is a consequence
of anisotropic scattering rates along $a$ and $b$ directions. Thus,
optical conductivity at low frequencies is the ideal tool to verify
our results. Unfortunately, as stated in Ref.\ \cite{Lucarelli11},
the experimental conditions in this temperature regime need to be
refined before definite conclusions can be drawn. Yet, it is encouraging
that the available data above $T_{N}$ on electron-doped samples indicate
a larger scattering rate along the \textbf{$b$ }direction \cite{Duzsa11,Lucarelli11},
in agreement with our results. We stress that our model is suitable
for the paramagnetic phase, where the sign of the resistivity anisotropy
is determined, according to the experimental results on various compounds
\cite{Chu10,Tanatar10,Blomberg11,Fisher11}. Below $T_{N}$, the spin
fluctuations associated with the magnetically ordered phase are naturally
anisotropic, and also give rise to anisotropic scattering. Both this
anisotropic scattering and the details of the anisotropic reconstruction
of the FS \cite{Calderon10,Fisher11,Kotliar11} will govern the temperature
evolution of $\Delta\rho$ below $T_{N}$. 

In summary, we showed that the onset of Ising-nematic order leads
to an anisotropy in the magnetic fluctuations associated with the
two magnetic ground states of the iron pnictides. As a result, the
same physics responsible for non-Fermi liquid behavior - the interference
of scattering by spin fluctuations and impurities \cite{Rosch99}
- also gives rise to anisotropic transport properties. The latter
are in agreement with several different experimental observations,
such as the opposite signs of $\Delta\rho$ in electron-doped and
hole-doped materials, as well as the suppression of $\Delta\rho$
upon sample annealing.

The authors thank J.-H. Chu, I. R. Fisher, R. Prozorov, and M. Tanatar
for useful discussions. Research at Ames Lab was supported by the
U.S. DOE, Office of BES, Materials Sciences and Engineering Division.

\begin{widetext}

\vspace{8 mm}

{\bf\Large \center Supplemental Material} \\

\renewcommand{\theequation}{S\arabic{equation}}
\setcounter{equation}{0}

The linearized Boltzmann equation determines the
electronic distribution function $f_{\mathbf{k},\lambda}$:

\begin{equation}
-e\mathbf{v}_{\mathbf{k},\lambda}\cdot\mathbf{n}\left(\frac{\partial f_{\mathbf{k},\lambda}^{0}}{\partial\xi_{\mathbf{k},\lambda}}\right)=\frac{1}{T}\sum_{\mathbf{k}',\lambda'}\left(\Phi_{\mathbf{k},\lambda}-\Phi_{\mathbf{k}',\lambda'}\right)f_{\mathbf{k}',\lambda'}^{0}\left(1-f_{\mathbf{k},\lambda}^{0}\right)t_{\mathbf{k},\mathbf{k}'}^{\lambda,\lambda'}\label{linearized_BE}\end{equation}

Here, $\Phi_{\mathbf{k},\lambda}$ is the non-equilibrium distribution
function, defined as $f_{\mathbf{k},\lambda}\equiv f_{\mathbf{k},\lambda}^{0}-\left(\frac{\partial f_{\mathbf{k},\lambda}^{0}}{\partial\varepsilon_{\mathbf{k},\lambda}}\right)\Phi_{\mathbf{k},\lambda}$,
with $f_{\mathbf{k},\lambda}^{0}$ denoting the Fermi-Dirac distribution
at energy $\varepsilon_{\mathbf{k},\lambda}$. $\mathbf{k}$ is the
momentum, $\lambda$ is the band index, $\mathbf{v}_{\mathbf{k},\lambda}=\partial\varepsilon_{\mathbf{k},\lambda}/\partial\mathbf{k}$,
and $\mathbf{n}$ is the unit vector parallel to the external electric
field. The collision term $t_{\mathbf{k},\mathbf{k}'}^{\lambda,\lambda'}$
comprises scattering by impurities and spin fluctuations \cite{Hlubina95,Rosch99}:

\begin{eqnarray}
t_{\mathrm{imp},\mathbf{k},\mathbf{k}'}^{\lambda,\lambda'} & = & \left(g_{\mathrm{imp}}^{2}\delta_{\lambda,\lambda'}+\bar{g}_{\mathrm{imp}}^{2}\left(1-\delta_{\lambda,\lambda'}\right)\right)\delta\left(\varepsilon_{\mathbf{k},\lambda}-\varepsilon_{\mathbf{k}',\lambda'}\right)\nonumber \\
t_{\mathrm{sf},\mathbf{k},\mathbf{k}'}^{\lambda,\lambda'} & = & \frac{2g_{\mathrm{sf}}^{2}}{\gamma}\: n\left(\varepsilon_{\mathbf{k},\lambda}-\varepsilon_{\mathbf{k}',\lambda'}\right)\mathrm{Im}\chi_{i}\left(\mathbf{k}-\mathbf{k}',\varepsilon_{\mathbf{k},\lambda}-\varepsilon_{\mathbf{k}',\lambda'}\right)\label{collision_terms}\end{eqnarray}
 with intraband impurity scattering amplitude $g_{\mathrm{imp}}^{2}\equiv\left|V_{\mathrm{imp}}^{\lambda=\lambda'}\right|^{2}$,
interband impurity scattering amplitude $\bar{g}_{\mathrm{imp}}^{2}=\left|V_{\mathrm{imp}}^{\lambda\neq\lambda'}\right|^{2}$,
and spin fluctuation scattering amplitude $g_{\mathrm{sf}}^{2}$.
Here, $n\left(\omega\right)$ is the Bose-Einstein distribution and
$\chi_{i}\left(\mathbf{q},\omega\right)$ is the dynamic susceptibility,
given by Eq.\ 2 of the paper. Since the electric field defines a
direction $\mathbf{n}$, it is convenient to write the non-equilibrium
distribution function as $\Phi_{\mathbf{k},\lambda}\equiv\Phi_{\mathbf{k},\lambda}^{\mu}n^{\mu}$,
with $\mu=1,2,3$ denoting Cartesian coordinates. Given $\Phi_{\mathbf{k},\lambda}^{\mu}$,
it is straightforward to obtain the conductivity tensor $\sigma^{\mu\nu}=-e\sum\limits _{\mathbf{k},\lambda}\left(\frac{\partial f_{\mathbf{k},\lambda}^{0}}{\partial\xi_{\mathbf{k},\lambda}}\right)v_{\mathbf{k},\lambda}^{\mu}\Phi_{\mathbf{k},\lambda}^{\nu}$.

To solve the Boltzmann equation (\ref{linearized_BE}), we write it
in terms of differential operators (see, for instance, J. M. Ziman, \emph{Electrons and phonons: the theory
of transport phenomena in solids}) $\hat{X}^{\mu}=\left(\hat{C}_{\mathrm{imp}}+\hat{C}_{\mathrm{sf}}\right)\hat{\Phi}^{\mu}$,
where $\hat{X}^{\mu}\equiv-e\, v_{\mathbf{k},\lambda}^{\mu}\left(\partial f_{\mathbf{k},\lambda}^{0}/\partial\varepsilon_{\mathbf{k},\lambda}\right)$
and $\hat{\Phi}^{\mu}$ are vectors ($\mu=1,2,3$), but also matrices
in the $\mathbf{k}\lambda$ space. $\hat{C}_{i}$ is defined by the
matrix elements:

\begin{equation}
\left(\hat{C}_{i}\hat{\Phi}^{\mu}\right)_{\mathbf{k}\lambda}=\frac{1}{T}\sum_{\mathbf{k}',\lambda'}\left(\Phi_{\mathbf{k},\lambda}^{\mu}-\Phi_{\mathbf{k}',\lambda'}^{\mu}\right)f_{\mathbf{k}',\lambda'}^{0}\left(1-f_{\mathbf{k},\lambda}^{0}\right)t_{\mathrm{i},\mathbf{k},\mathbf{k}'}^{\lambda,\lambda'}\label{aux_matrix_C}\end{equation}

In this language, it is convenient to introduce the scalar product
$\left\langle \left.\hat{X}^{\mu}\right|\left.\hat{Y}^{\nu}\right.\right\rangle =\sum_{\mathbf{k},\lambda}X_{\mathbf{k},\lambda}^{\mu}Y_{\mathbf{k},\lambda}^{\nu}$,
from which $\sigma^{\mu\nu}=\left\langle \left.\hat{X}^{\mu}\right|\hat{\Phi}^{\nu}\right\rangle $.
We consider that impurities are the dominant scatterers, while scattering
by spin fluctuations perturbatively suppresses the conductivity. This
is always true at low enough temperatures, where $g_{\mathrm{sf}}^{2}\left(T/\gamma\right)$
is small compared to $g_{\mathrm{imp}}^{2}$. To lowest order, the
solution of the Boltzmann equation is $\hat{\Phi}^{\mu}=\hat{\Phi}^{0,\mu}+\delta\hat{\Phi}^{\mu}$,
where $\hat{\Phi}^{0,\mu}$ is the solution in the presence of impurities
only and $\delta\hat{\Phi}^{\mu}$ is the solution of $\hat{C}_{\mathrm{imp}}\delta\hat{\Phi}^{\mu}=-\hat{C}_{\mathrm{sf}}\hat{\Phi}^{0,\mu}$.
Therefore, the conductivity becomes:

\begin{equation}
\sigma^{\mu\mu}=\left\langle \hat{\Phi}^{0,\mu}\left|\hat{C}_{\mathrm{imp}}\right|\hat{\Phi}^{0,\mu}\right\rangle -\left\langle \hat{\Phi}^{0,\mu}\left|\hat{C}_{\mathrm{sf}}\right|\hat{\Phi}^{0,\mu}\right\rangle \label{perturbation_2}\end{equation}

The first term is the constant impurity contribution to the conductivity,
$\sigma_{\mathrm{imp}}$, whereas the second term is the temperature
dependent contribution from spin fluctuations (see Eq.\ 1 of the
main text). The calculation of $\hat{\Phi}^{0,\mu}$, which satisfy
$\hat{X}^{\mu}=\hat{C}_{\mathrm{imp}}\hat{\Phi}^{0,\mu}$, is tedious
but straightforward. We use the variational principle, as outlined
in \cite{Rosch99}, and the three-band model discussed in the
paper, and obtain, for the electric field along $\hat{x}$:

\begin{eqnarray}
\Phi_{1,\theta}^{0,x} & = & -\frac{ev_{F}}{\nu\left(g_{\mathrm{imp}}^{2}+2\bar{g}_{\mathrm{imp}}^{2}\right)}\cos\theta\nonumber \\
\Phi_{(2,3),\theta}^{0,x} & = & \frac{ev_{F}}{\nu\left(g_{\mathrm{imp}}^{2}+2\bar{g}_{\mathrm{imp}}^{2}\right)}\left(\frac{1+\bar{\delta}_{0}\mp\bar{\delta}_{2}\pm2\bar{\delta}_{2}\cos2\theta}{\sqrt{1+\bar{\delta}_{0}\pm\bar{\delta}_{2}\cos2\theta}}\right)\cos\theta\label{phi_0_x}\end{eqnarray}

Here, $\theta$ is the polar angle and $v_{F}$, $\varepsilon_{F}$,
and $\nu$ are, respectively, the Fermi velocity, Fermi energy, and
density of states of the hole pocket. $\bar{\delta}_{i}=\delta_{i}/\varepsilon_{F}$
depends only on the band structure parameters and is given by:

\begin{eqnarray}
\frac{\delta_{0}}{\varepsilon_{F}} & = & 2\bar{\mu}+\left(1-\frac{\bar{m}_{x}+\bar{m}_{y}}{2\bar{m}_{x}\bar{m}_{y}}\right)\nonumber \\
\frac{\delta_{2}}{\varepsilon_{F}} & = & \frac{\bar{m}_{x}-\bar{m}_{y}}{2\bar{m}_{x}\bar{m}_{y}}\label{aux_band_structure}\end{eqnarray}
 with $\bar{m}_{i}=m_{i}/m$ and $\bar{\mu}=\mu/\varepsilon_{F}$.
The residual conductivity due to impurities is:

\begin{equation}
\sigma_{\mathrm{imp}}=\frac{e^{2}v_{F}^{2}}{2\left(g_{\mathrm{imp}}^{2}+2\bar{g}_{\mathrm{imp}}^{2}\right)}\left[1+4\left(1+\bar{\delta}_{0}\right)-2\sqrt{\left(1+\bar{\delta}_{0}\right)^{2}-\bar{\delta}_{2}^{2}}\right]\label{resistivity_anisotropy}\end{equation}

For an electric field parallel to the $\hat{y}$ direction, we can
use the transformation properties of the non-equilibrium distribution
functions under a $\pi/2$ rotation: $\Phi_{1,\theta}^{0,y}=\Phi_{1,\tilde{\theta}}^{0,x}$,
$\Phi_{2,\theta}^{0,y}=\Phi_{3,\tilde{\theta}}^{0,x}$, and $\Phi_{3,\theta}^{0,y}=\Phi_{2,\tilde{\theta}}^{0,x}$,
with $\tilde{\theta}=\theta-\pi/2$.

We can now evaluate the matrix elements $\left\langle \hat{\Phi}^{0,\mu}\left|\hat{C}_{\mathrm{sf}}\right|\hat{\Phi}^{0,\mu}\right\rangle $
using the solutions in Eqs. \ref{phi_0_x}. After defining:

\begin{eqnarray}
\omega_{\theta,\theta'}^{(i)} & = & \left(\xi_{i}/a_{0}\right)^{-2}+\left(2+\bar{\delta}_{0}\pm\bar{\delta}_{2}\cos2\theta'\right)-2\cos\left(\theta-\theta'\right)\sqrt{1+\bar{\delta}_{0}\pm\bar{\delta}_{2}\cos2\theta'}\nonumber \\
 &  & \pm\eta\left[\cos2\theta-2\cos\left(\theta+\theta'\right)\sqrt{1+\bar{\delta}_{0}\pm\bar{\delta}_{2}\cos2\theta'}+\cos2\theta'\left(1+\bar{\delta}_{0}\pm\bar{\delta}_{2}\cos2\theta'\right)\right]\label{omega_phi}\end{eqnarray}
 we obtain (hereafter we omit the superscript $0$ in $\Phi_{1,\theta}^{0,\mu}$
for convenience):

\begin{equation}
\frac{\rho^{\mu\mu}-\rho_{\mathrm{imp}}}{\rho_{\mathrm{imp}}}=\zeta t\sum_{i=2,3}\int d\theta\, d\theta'\left(\Phi_{1,\theta}^{\mu}-\Phi_{i,\theta'}^{\ \mu}\right)^{2}\:\frac{\sqrt{\left(\omega_{\theta,\theta'}^{(i)}+\frac{2\pi}{3}t\right)\left(\omega_{\theta,\theta'}^{(i)}+\frac{2\pi}{3}t+\eta_{z}^{2}\right)}-\sqrt{\omega_{\theta,\theta'}^{(i)}\left(\omega_{\theta,\theta'}^{(i)}+\eta_{z}^{2}\right)}}{2\pi\sqrt{\omega_{\theta,\theta'}^{(i)}\left(\omega_{\theta,\theta'}^{(i)}+\eta_{z}^{2}\right)\left(\omega_{\theta,\theta'}^{(i)}+\frac{2\pi}{3}t\right)\left(\omega_{\theta,\theta'}^{(i)}+\frac{2\pi}{3}t+\eta_{z}^{2}\right)}}\label{resistivity_anisotropy}\end{equation}
 with $\rho_{\mathrm{imp}}=\sigma_{\mathrm{imp}}^{-1}$, $t=T/\gamma\left(k_{F}a_{0}\right)^{2}$
and the pre-factor:

\begin{equation}
\zeta=\frac{2g_{\mathrm{sf}}^{2}}{g_{\mathrm{imp}}^{2}+2\bar{g}_{\mathrm{imp}}^{2}}\left[1+4\left(1+\bar{\delta}_{0}\right)-2\sqrt{\left(1+\bar{\delta}_{0}\right)^{2}-\bar{\delta}_{2}^{2}}\right]^{-1}\label{aux_pre_factor}\end{equation}

In Fig.\ 3 of the paper, we show the numerical solution of Eq.\ (\ref{resistivity_anisotropy})
for the parameters discussed in the main text, using $\xi$ {[}appearing
in Eq.\ (S8)] that comes from the solution of the self-consistent
equations (3) of the main text.

At low temperatures $t\ll1$, we can obtain an analytical solution
for the resistivity anisotropy. We consider that the system is in
the orthorhombic paramagnetic phase, where the correlation length
associated with each ordering vector is given by $\xi_{i}^{-2}=\xi^{-2}\mp\varphi$,
where $\varphi>0$ is the Ising-nematic order parameter. To simplify
our analysis, we consider a magnetic three-dimensional QCP. However,
our results are more general and hold even away from the QCP.

For the hard mode (short correlation length), $\xi_{3}^{-2}=\xi^{-2}+\varphi$,
the $t\ll1$ expansion of Eq.\ (\ref{resistivity_anisotropy}) yields:

\begin{equation}
\left(\frac{\rho^{\mu\mu}-\rho_{\mathrm{imp}}}{\rho_{\mathrm{imp}}}\right)_{\mathrm{hard}}=t^{2}\left\{ \zeta\int d\theta\, d\theta'\:\frac{\left(\Phi_{1,\theta}^{\mu}-\Phi_{3,\theta'}^{\ \mu}\right)^{2}\left(2\varphi+f_{\theta,\theta'}^{(3)}+\frac{\eta_{z}^{2}}{2}\right)}{3\left[\left(2\varphi+f_{\theta,\theta'}^{(3)}\right)\left(2\varphi+f_{\theta,\theta'}^{(3)}+\eta_{z}^{2}\right)\right]^{3/2}}\right\} \label{hard_mode}\end{equation}
 where we defined the auxiliary function $f_{\theta,\theta'}^{(i)}=\omega_{\theta,\theta'}^{(i)}-\left(\xi_{i}/a_{0}\right)^{-2}$
{[}see Eq.\ (\ref{omega_phi})]. Thus, the contribution of the hard
spin fluctuations to the resistivity vanishes as $t^{2}$. For the
soft mode $\xi_{2}^{-2}=\xi^{-2}-\varphi$ the situation is different,
since $\xi_{2}^{-2}\rightarrow0$ as $t\rightarrow0$. Then, one has
to divide the Fermi surface into two regions \cite{Rosch99}: one
around the hot spots $\theta_{hs}$, with width $t$ ($f_{\theta,\theta'}^{(2)}\sim t$),
and another with $f_{\theta,\theta'}^{(2)}\gg t$, called the {}``cold
region''. Repeating the same steps that led to Eq.\ (\ref{hard_mode}),
we find that the contribution from the cold region also vanishes as
$t^{2}$. Near the hot spots $\theta_{hs}$, an expansion similar
to the one outlined in \cite{Rosch99} yields:\begin{equation}
\left(\frac{\rho^{\mu\mu}-\rho_{\mathrm{imp}}}{\rho_{\mathrm{imp}}}\right)_{\mathrm{soft}}=\left(\frac{4\sqrt{2\pi}\zeta c}{\sqrt{3}\eta_{z}}\right)\: t^{3/2}\left(\Phi_{1,\theta_{hs}}^{\mu}-\Phi_{2,\theta_{hs}}^{\ \mu}\right)^{2}\label{soft_mode}\end{equation}
 which is the Eq.\ (4) of the paper. Here, $c>0$ is a constant coming
from the Jacobian of appropriate coordinate transformations. Since
the contribution from the soft mode to the resistivity vanishes as
$t^{3/2}$, while the other contributions vanish as $t^{2}$, the
former dominates the transport at low temperatures. Therefore, the
resistivity anisotropy at $t\ll1$ is:

\begin{equation}
\frac{\rho^{yy}-\rho^{xx}}{\rho_{\mathrm{imp}}}=\kappa T^{3/2}\left[\left(\Phi_{1,\theta_{hs}}^{y}-\Phi_{2,\theta_{hs}}^{\ y}\right)^{2}-\left(\Phi_{1,\theta_{hs}}^{x}-\Phi_{2,\theta_{hs}}^{\ x}\right)^{2}\right]\label{aux_low_T}\end{equation}
 with $\kappa>0$. Using Eqs.\ (\ref{phi_0_x}) and the fact that
the hot spots are at $\theta_{hs}=\arccos(\delta_{0}/\delta_{2})/2$,
we obtain:

\begin{equation}
\frac{\rho^{yy}-\rho^{xx}}{\rho_{\mathrm{imp}}}=\kappa T^{3/2}\left[\frac{\bar{\delta}_{0}\left(2-\bar{\delta}_{0}\right)^{2}}{\bar{\delta}_{2}}+\left(4-\bar{\delta}_{0}\right)\bar{\delta}_{2}\right]\label{low_T}\end{equation}

Thus, for a fixed value of the mass anisotropy ($\delta_{2}$), the
sign of the resistivity anisotropy changes from positive to negative
if the chemical potential ($\delta_{0}$) is negative and large enough.
The {}``phase diagram'' in Fig.\ 4 of the paper is obtained from
this expression.

\end{widetext}


\begin{thebibliography}{10}
\bibitem{Kivelson98} S. A. Kivelson, E. Fradkin, and V. J. Emery,
Nature \textbf{393}, 550 (1998).

\bibitem{Chu10} J. Chu \emph{et al}., Science \textbf{329}, 824 (2010).

\bibitem{Tanatar10} M. A. Tanatar \emph{et al}., Phys. Rev. B \textbf{81},
184508 (2010).

\bibitem{Shen11} M. Yi \emph{et al}., arXiv:1011.0050 (2010).

\bibitem{Davis10} T.-M. Chuang \emph{et al.,} Science \textbf{327},
181 (2010).

\bibitem{Duzsa11} A. Dusza \emph{et al}., Europhys. Lett. \textbf{93},
37002 (2011).

\bibitem{Lucarelli11} A. Lucarelli \emph{et al}., arXiv:1107.0670.

\bibitem{Uchida11} M. Nakajima \emph{et al}., arXiv:1106.4967.

\bibitem{Si08} Q. Si and E. Abrahams, Phys. Rev. Lett. \textbf{101},
076401 (2008).

\bibitem{Yildirim08} T. Yildirim, Phys. Rev. Lett. \textbf{101},
057010 (2008).

\bibitem{Fang08} C. Fang \emph{et al}., Phys. Rev. B \textbf{77,}
224509 (2008).

\bibitem{Xu08} C. Xu, M. Müller, and S. Sachdev, Phys. Rev. B \textbf{78},
020501(R) (2008).

\bibitem{FernandesPRL10} R. M. Fernandes \emph{et al}., Phys. Rev.
Lett. \textbf{105}, 157003 (2010).

\bibitem{Eremin10} I. Eremin and A. V. Chubukov, Phys. Rev. B \textbf{81},
024511 (2010).

\bibitem{Antropov11} A. L. Wysocki, K. D. Belashchenko, and V. P.
Antropov, Nat. Phys. (2011).

\bibitem{chandra} P. Chandra, P. Coleman, and A.I. Larkin, Phys.
Rev. Lett. \textbf{64}, 88 (1990).

\bibitem{Nandi09} S. Nandi \emph{et al}., Phys. Rev. Lett. \textbf{104},
057006 (2010).

\bibitem{FernandesPRB10} R. M. Fernandes \emph{et al}., Phys. Rev.
B \textbf{81}, 140501(R) (2010).

\bibitem{Chen10} C. C. Chen \emph{et al}., Phys. Rev. B \textbf{82},
100504 (2010).

\bibitem{Calderon10} B. Valenzuela, E. Bascones, and M. J. Calderón,
Phys. Rev. Lett. \textbf{105}, 207202 (2010).

\bibitem{Blomberg11} E. C. Blomberg \emph{et al}., arXiv:1101.0274.

\bibitem{Ying10} J. J. Ying \emph{et al}., arXiv:1012.2731.

\bibitem{Tanatar_private} M. Tanatar and R. Prozorov, private communication.

\bibitem{Hlubina95} R. Hlubina and T. M. Rice, Phys. Rev. B \textbf{51},
9253 (1995).

\bibitem{Rosch99} A. Rosch, Phys. Rev. Lett. \textbf{82}, 4280 (1999).

\bibitem{Kemper11} A. F. Kemper \emph{et al.}, Phys. Rev. B \textbf{83},
184516 (2011) .

\bibitem{Zhang10} J. Zhang, R. Sknepnek, and J. Schmalian, Phys.
Rev. B \textbf{82}, 134527 (2010).

\bibitem{Diallo10} S. O. Diallo \emph{et al}., Phys. Rev. B \textbf{81},
214407 (2010); D. S. Inosov \emph{et al}., Nat. Phys. \textbf{6},
178 (2010).

\bibitem{Kasahara10} S. Kasahara \emph{et al}., Phys. Rev. B \textbf{81,}
184519 (2010).

\bibitem{Abrahams09} J. Dai, Q. Si, J.-X. Zhu, and E. Abrahams, PNAS
\textbf{106}, 4118 (2009).

\bibitem{Liu10} C. Liu \emph{et al}., Nat. Phys. \textbf{6}, 419
(2010).

\bibitem{Westfahl98} R. M. Fernandes, J. Schmalian, and H. Westfahl
Jr, Phys. Rev. B \textbf{78}, 184201 (2008). 

\bibitem{Fisher11} H. H. Kuo \emph{et al}., arXiv:1103.4535.

\bibitem{Kotliar11} Z. P. Yin, K. Haule, and G. Kotliar, Nat. Phys.
\textbf{7}, 294 (2011).
\end{thebibliography}
\end{document}